\AtBeginDocument{\RenewCommandCopy\qty\SI}

\documentclass[
 aip,
 amsmath,
 amssymb,
 preprint
]{revtex4-1}

\usepackage{graphicx}
\usepackage{dcolumn}
\usepackage{bm}

\usepackage[utf8]{inputenc}
\usepackage[T1]{fontenc}
\usepackage{mathptmx}
\usepackage{etoolbox}

\usepackage{siunitx} 
\usepackage{braket} 
\usepackage{amssymb} 
\usepackage{comment} 

\usepackage[dvipsnames]{xcolor} %
\usepackage{booktabs} 
\usepackage{amsmath} 
\usepackage{physics} 
\usepackage[colorlinks=true, linkcolor=black, citecolor=black, urlcolor=blue]{hyperref}
\usepackage[capitalise]{cleveref} 

\makeatletter
\def\@email#1#2{%
 \endgroup
 \patchcmd{\titleblock@produce}
  {\frontmatter@RRAPformat}
  {\frontmatter@RRAPformat{\produce@RRAP{*#1\href{mailto:#2}{#2}}}\frontmatter@RRAPformat}
  {}{}
}%
\makeatother

\preprint{AIP/123-QED}

\newcommand*\Bnv {B_\mathrm{NV}} 
\newcommand*\Boff {B_\mathrm{\perp}} 
\newcommand*\Bnvgrad {G_\mathrm{z}} 
\newcommand*\phnv {\phi_\mathrm{NV}} 
\newcommand*\thnv {\theta_\mathrm{NV}} 
\newcommand*\mum {m_\mathrm{dip}} 
\newcommand*\vecmum {\vb{m}_\mathrm{dip}} 
\newcommand*\Bb {B_\mathrm{bias}} 
\newcommand*\Br {B_\mathrm{r}} 
\newcommand*\Ms {M_\mathrm{sat}} 
\newcommand*\Cz {C_\mathrm{z}} 
\newcommand*\Cgen {C_\mathrm{z,gen}} 
\newcommand*\Bdip {\vb{B}_\mathrm{dip}} 
\newcommand*\rzz {r_\mathrm{z,0}} 
\newcommand*\zstep {z_\mathrm{step}} 
\newcommand*\mad {\sigma} 
\newcommand*\dnv {d_\mathrm{NV}} 
\newcommand*\nnv {\hat{\vb{n}}_\mathrm{NV}} 

\DeclareSIUnit{\NA}{NA}

\newcommand{\teresa}[1]{{\color{red} \bf Teresa: #1}}

\begin{document}
\title{Measuring high field gradients of cobalt nanomagnets in a spin-mechanical setup}

\author{Felix Hahne}
\thanks{These authors contributed equally to this work.}
\affiliation{Niels Bohr
Institute, University of Copenhagen, Blegdamsvej 17, 2100, Copenhagen, Denmark}
\affiliation{Center for Hybrid Quantum Networks, Niels Bohr Institute, University of Copenhagen, Blegdamsvej 17, 2100, Copenhagen, Denmark}

\author{Teresa Klara Pfau}
\thanks{These authors contributed equally to this work.}
 \affiliation{Niels Bohr
Institute, University of Copenhagen, Blegdamsvej 17, 2100, Copenhagen, Denmark}
\affiliation{Center for Hybrid Quantum Networks, Niels Bohr Institute, University of Copenhagen, Blegdamsvej 17, 2100, Copenhagen, Denmark}

\author{Liza \v{Z}aper}
\thanks{These authors contributed equally to this work.}
 \affiliation{Department of Physics, University of Basel, Klingelbergstrasse 82, 4056 Basel, Switzerland}

\author{Lucio Stefan}
 \affiliation{Niels Bohr
Institute, University of Copenhagen, Blegdamsvej 17, 2100, Copenhagen, Denmark}
\affiliation{Center for Hybrid Quantum Networks, Niels Bohr Institute, University of Copenhagen, Blegdamsvej 17, 2100, Copenhagen, Denmark}

\author{Thibault Capelle}
\affiliation{Niels Bohr
Institute, University of Copenhagen, Blegdamsvej 17, 2100, Copenhagen, Denmark}
\affiliation{Center for Hybrid Quantum Networks, Niels Bohr Institute, University of Copenhagen, Blegdamsvej 17, 2100, Copenhagen, Denmark}

\author{Andrea Ranfagni}
 \affiliation{Niels Bohr
Institute, University of Copenhagen, Blegdamsvej 17, 2100, Copenhagen, Denmark}
\affiliation{Center for Hybrid Quantum Networks, Niels Bohr Institute, University of Copenhagen, Blegdamsvej 17, 2100, Copenhagen, Denmark}

  \author{ Martino Poggio}
 \affiliation{Department of Physics, University of Basel, Klingelbergstrasse 82, 4056 Basel, Switzerland} 

  \author{Albert Schliesser}
\affiliation{Niels Bohr
Institute, University of Copenhagen, Blegdamsvej 17, 2100, Copenhagen, Denmark}
\affiliation{Center for Hybrid Quantum Networks, Niels Bohr Institute, University of Copenhagen, Blegdamsvej 17, 2100, Copenhagen, Denmark}

\date{\today}

\begin{abstract}
Hybrid systems composed of a single nitrogen-vacancy center spin magnetically coupled
to a macroscopic mechanical resonator constitute promising platforms for the realization
of quantum information protocols and for quantum sensing applications.
The magnetic structure that mediates the interaction must ensure high field gradients while
preserving the spin and mechanical properties.
We present a spin-mechanical setup built around a cobalt nanomagnet grown with focused electron beam-induced deposition.
The magnetic structure is fully characterized, and a maximum gradient of $\SI{170}{\kilo\tesla\per\meter}$ is directly
measured at a spin-oscillator distance of a few hundred nanometers.
Spin coherence was preserved at the value of $\SI{20}{\micro\second}$ up to a gradient of $\SI{25}{\kilo\tesla\per\meter}$.
The effect of the mechanical motion onto the spin dynamics was observed, thus signifying the presence of spin-mechanics coupling. 
Given the noninvasive nature of the nanomagnet deposition process,
we foresee the adoption of such structures in hybrid platforms with high-quality factor resonators, in the "magnet on oscillator" configuration.

\end{abstract}

\maketitle

Following the landmark experiment \cite{Rugar2004} that revealed how a single electron spin can exert a measurable force on a macroscopic cantilever, various hybrid platforms composed of two-level systems coupled to macroscopic oscillators have emerged by integrating mechanical resonators with
superconducting qubits \cite{LaHaye2009,O’Connell2010,Pirkkalainen2013,Etaki2008}
and solid state defects \cite{Montinaro2014,Golter2016}.
Thanks to the mature technology of ultra-high quality factor oscillators 
 \cite{Bachtold2022}, these platforms are appealing as testbeds for quantum 
information processing experiments \cite{Andersen2015} and also for pushing the 
boundaries of quantum sensing \cite{Eichler2022}. 

The paradigmatic system of a single spin coupled to a mechanical resonator
device has been successfully implemented using NV defects  in
diamond \cite{Doherty2013} coupled to the mechanics via strain \cite{Barfuss2015strong,Ovartchaiyapong2014} or magnetically \cite{Arcizet2011,kolkowitz2012coherent,Pigeau2015,Oeckinghaus2020,Fung2024,Hong2012,Gieseler2020}. 

When a magnetically-decorated oscillator is placed in the vicinity
of an NV center \cite{Rabl2009a},
the Zeeman interaction results in a parametric single-phonon coupling strength $g_0 = 2 \pi \gamma_\mathrm{NV}  G_\mathrm{z} 
z_\mathrm{zpf}$,
where $\gamma_{\mathrm{NV}} \simeq  \SI{28}{\giga\hertz\per\tesla}$
is the gyromagnetic ratio of the NV center, $G_\mathrm{z} = \partial \Bnv / \partial z$ is the gradient at the NV rest position---along the oscillator’s motion direction ($z$)---of the magnetic field component ($\Bnv$) parallel to the NV spin quantization axis, and $z_{\mathrm{zpf}} = \sqrt{\hbar / 2m\Omega_\mathrm{m}}$ is the zero-point motion amplitude, where $\hbar$ is the reduced Planck constant, $\Omega_\mathrm{m}$ is the oscillator’s angular frequency, and $m$ is its mass.

A crucial requirement for implementing quantum protocols
is achieving a quantum cooperativity greater than one, with the cooperativity defined\cite{kolkowitz2012coherent} as $C=g_0^2 T_2 \Gamma^{-1} $, 
where $\Gamma$ in the decoherence rate of the oscillator, and $T_2$ denotes the spin transverse relaxation time under a specific decoupling sequence.
When $C>1$, the zero-point mechanical fluctuations could be 
resolved within a few shot measurements \cite{kolkowitz2012coherent,
Bennett2012}. In this regime, a hot mechanical resonator could be exploited to
realize quantum gates between two spins in a hybrid nano-electromechanical system \cite{Rabl2010}, or -- when $C>4$ -- to efficiently generate entanglement between two spin qubits
\cite{Rosenfeld2021}.

An important challenge currently faced by 
the community in this regard, is the research of a suitable magnetic structure, which must achieve high 
magnetic field gradients while preserving both the oscillator’s 
properties and the overall system coherence \cite{Lee2017}.
Different methods have been explored, 
including the 
use of NdFeB magnetic structures \cite{Arcizet2011}, a CoFe magnetic film evaporated onto a 
quartz tip \cite{Hong2012}, a CoCr layer deposited on a tip \cite{kolkowitz2012coherent}, and NdFeB beads \cite{Fung2024,Pigeau2015}.

Here, we harness the nanometric control of focused electron beam 
induced deposition (FEBID) \cite{DeTeresa2016}  to grow soft cobalt nanomagnets 
on a silicon chip. The nanomagnetic structure, which exhibits strong magnetic gradients, is installed in a spin-mechanics setup based on a scanning NV microscopy configuration. 
\begin{figure*}[t!]
  \centering
  \includegraphics[width=0.95\textwidth]{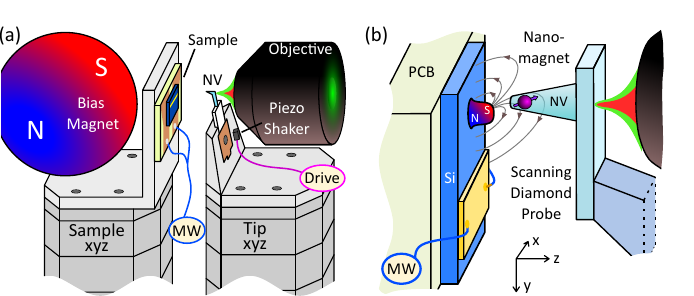}
  \caption{\small{
  Schematic of the scanning setup. 
  (a) Sample and NV probe are mounted on two $x,y,z$ piezo scanners. 
  The nanomagnets are magnetized by a $\SI{4}{\centi\meter}$-diameter spherical permanent magnet at a distance of a few millimeters.
  A $\SI{0.9}{\NA}$ and $\SI{1}{\milli\meter}$ working distance objective focuses the collimated excitation beam onto the NV attached to a quartz tuning fork.
  A piezo shaker mechanically excites the tuning fork motion.
  (b) Depicted is a single nanomagnet in proximity of the microwave (MW) stripline and the NV probe in the laboratory frame.
  The silicon chip is placed in the apparatus with a printed circuit board holder that also provides the microwave contacts.
  }}
  \label{setup}
\end{figure*}
The scanning NV microscopy configuration we use for this work utilizes $x,y,z$ piezoelectric nanopositioners (attocube), allowing nanometer-resolution spatial positioning of the mounted tuning fork with the NV center relative to a selected sample with the nanomagnets (\cref{setup}(a)).
The nanomagnets are magnetized by a homogeneous external magnetic bias field of up to $\SI{140}{\milli\tesla}$ generated by a spherical magnet which is aligned to the NV quantization axis.
Excitation with laser light at a wavelength of $\SI{515}{nm}$ and readout of the NV photoluminescence rate are carried out using a home-built confocal microscope with a $\SI{0.9}{\NA}$ objective (Olympus MPlanFL N 100x).
The scanning system is configured vertically to allow the insertion of a permanent magnet for magnetic biasing on the backside of the sample chip.

The core of the system consists of a {100}-cut diamond scanning probe (QZabre QST \cite{qzabre}), hosting the NV center, and a Si chip with the Co nanomagnets plus an integrated microwave stripline for spin control (\cref{setup}(b)).
The nanomagnet of this study is approximately $\SI{40}{\micro\metre}$ away from the stripline and has a cylindrical shape with $\SI{830}{\nano\metre}$  height and $\SI{400}{\nano\metre}$ diameter.

To grow these nanomagnets we employ FEBID with the geometry  defined as a circular pattern and the following
parameters: an acceleration voltage of 5 kV, a beam current of
\SI{100}{\pico\ampere}, a dwell time of \SI{1}{\micro\second}, and a precursor flux corresponding to a vacuum pressure varying
in the range \SIrange{1.1e-6}{1.2e-6}{\milli\bar} (see SI \cref{suppl:nanomagnet_deposition}).

\begin{figure*}[t!]
  \includegraphics[width=0.99\textwidth]{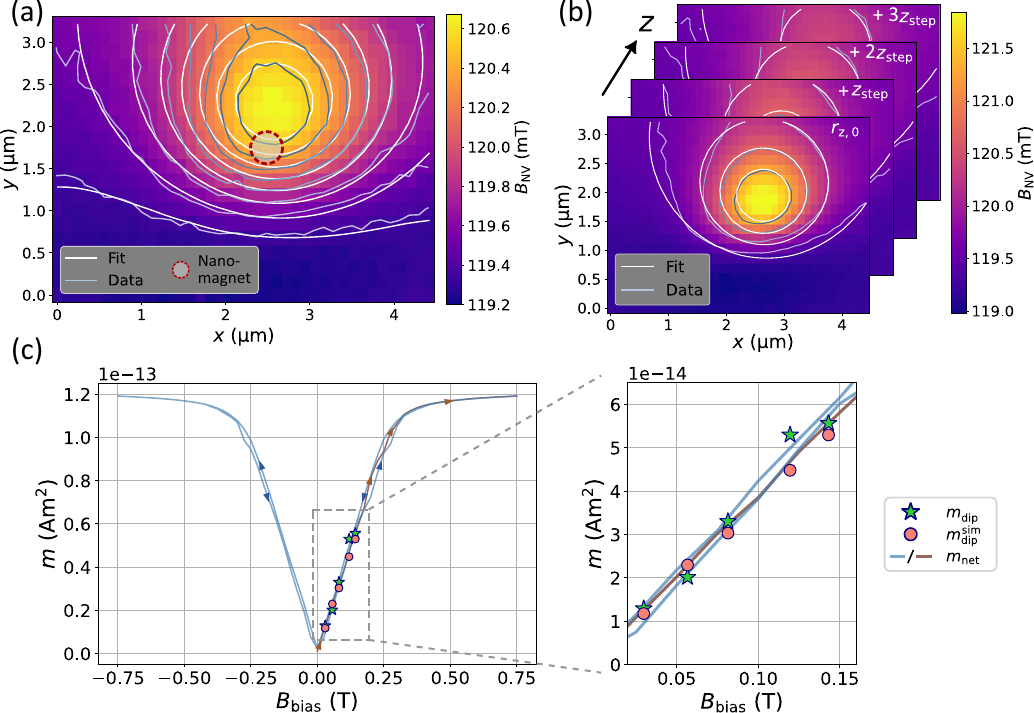}
  \caption{
  Dipole fitting and stray field characterization.
  (a) A map of the measured $\Bnv$ at an estimated height of $h = \SI{1294(40)}{\nano\meter}$ above the nanomagnet and with $\Bb \sim \SI{120}{\milli\tesla}$.
  The white contour lines from a dipole model fit agree with the gray data contour lines of equal value. 
  The indicated nanomagnet is centered around the dipole $(x,y)$ position.
  (b) Simultaneous dipole fitting of four stray-field maps acquired at successive heights above the sample in increments of $\zstep = \qty{166(5)}{nm}$.
  A global dipole depth $\rzz$ is enforced across all maps, while the relative height offset is fixed by the known step size.
  (c) Plot of the fitted dipole magnetic moment $\mum$ (green stars), dipole fitted $\mum^\mathrm{sim}$ (red dots) and simulated net magnetization $m_\mathrm{net}$ (blue and brown lines) with increasing external biasing field.
  For the hysteresis curve of $m_\mathrm{net}$ the external field is varied from $\qty{0}{mT}\rightarrow\qty{750}{mT}\rightarrow\qty{-750}{mT}\rightarrow\qty{750}{mT}$ in increments of $\qty{25}{mT}$.
  The brown line marks the initial upsweep from $\qty{0}{mT}$ to $\qty{750}{mT}$, while the blue line corresponds to the remainder of the sweep.
  Arrows on the lines indicate the sweep direction of the external field.
  The inset displays a zoomed-in section of the boxed data region for improved visibility of the individual points.
  }
  \label{dipolefit}
\end{figure*}

We perform magnetometry measurements\cite{Balasubramanian2008,Maletinsky2012,Rondin2014} to characterize the stray field $\vb{B}_\mathrm{nm}(x,y,z)$ emanating from the nanomagnet. We measure the shift in the lower energy electron spin resonance of the optical ground state spin triplet while performing a $xy$ raster scans on planes above the magnetic structure.
The optically detected magnetic resonance (ODMR) is performed at each position $(x,y)$ by applying 
a continuous-wave microwave sweep and acquiring the NV center photoluminescence rate\cite{Gruber1997}.
The scans are performed on a single nanomagnet under ambient conditions, without feedback and hovering at a fixed height above the substrate.
The spin resonance frequencies $\nu_\mathrm{res}^{\pm}$, in the relevant experimental configuration, reads:
\begin{equation}
    \nu_\mathrm{res}^\pm \simeq \abs{\nu_0 \pm  \gamma_\mathrm{NV} \Bnv} \,. 
    \label{eq:Zeeman interaction}
\end{equation}
Here, $\nu_0 \simeq 2.87\ \mathrm{GHz}$  is the ground state zero-field splitting parameter and the total magnetic field is defined as $\Bnv = \vb{B}_\mathrm{nm}\cdot\nnv + \Bb$, where $\nnv$ is the spin
quantization axis direction, and the bias field is assumed to be aligned to the spin axis.
We are neglecting the local off-axis strain field contribution to the energy levels, and we assume to operate the system in the regime where $\Boff \ll \abs{ \nu_0 /\gamma_\mathrm{NV} - \abs{\Bnv}}$, where $\Boff$ is the magnitude of the field component orthogonal to the spin quantization axis.
All acquired scanned maps display stray field patterns consistent with a magnetic dipole field as confirmed by dipole model fits shown in \cref{dipolefit}(a).

Approximating the nanomagnet as a dipole, we can extract the magnitude of the magnetic dipole moment $\mum$ from the dipole model fits (see SI \cref{suppl: dipole fit}).
We acquire a set of four scanned maps at different scanning heights to capture the stray field distribution at multiple distances from the nanomagnet.
We increment the heights from $\qtyrange{960}{1460}{\nano\metre}$ above the nanomagnet in steps of $\zstep = \SI{166(5)}{\nano\metre}$.
The maps of a set share the same experimental conditions otherwise.
A simultaneous fit across all four scanned maps as shown in \cref{dipolefit}(b), incorporating the known height steps $\zstep$, 
additionally constrains the fit parameters $\rzz$ (depth of the approximated point-like dipole within the nanomagnet) and $\mum$, which for individual maps can float in a correlated manner. 
The fit then allows quantitative inference 
of the magnetic moment magnitude $\mum$.
We repeat this fit procedure for external bias fields $\Bb$ between from \qtyrange{30}{140}{\milli\tesla} and plot $\mum(\Bb)$ in \cref{dipolefit}(c,d) (green stars).
The estimated dipole moment magnitude $\mum(\Bb)$ increases linearly, indicating a soft magnetic nature of the FEBID grown nanomagnet as expected from their typically amorphous structure \cite{Fernandez2020}.

For comparison, we similarly perform the simultaneous fits on stray field maps generated by a simulation of the nanomagnet, that is based on the dipole fit and AFM topography (see SI \cref{suppl:micromatnetic simulation}), extracting $\mum^{\mathrm{sim}}(\Bb)$ (red circles in \cref{dipolefit}(c,d)) that agrees well with $\mum(\Bb)$.
Furthermore, we can calculate the net magnetic moment magnitude of the simulated nanomagnet via the volume integral of the magnetization $m_\mathrm{net} \approx |\sum_{ijk} \vb{M}_{ijk}\Delta V|$, where $\vb{M}_{ijk}$ is the magnetization in each simulation cell of volume $\Delta V$.
The moment magnitude $m_\mathrm{net}(\Bb)$ is depicted in \cref{dipolefit}(c,d).
The agreement between $\mum$, $\mum^{\mathrm{sim}}$, and $m_\mathrm{net}$ demonstrates mutual validation across of the experimental measurements, the dipole model fit, and the simulation.

The simulation of the magnetic structure predicts that the nanomagnet---that is $\mum(\Bb)$---approaches saturation \cite{Coey2010} at around $\Bb\sim\SI{300}{\milli\tesla}$.
Supporting this prediction we find agreement between the ratio of the bias fields $\SI{140}{mT}/\SI{300}{\milli\tesla} = 0.47$ and the magnetization ratio $M/\Ms = 0.45$, where $M = \mum(\Bb = \SI{140}{\milli\tesla})/V$ with the nanomagnet volume $V$ and saturation magnetization $\Ms = \SI{1.26}{\mega\ampere\per\meter}$ used for the simulation \cite{doi:10.1021/acsanm.3c05470}.

\begin{figure*}[t!]
\center
  \includegraphics[width=0.99\linewidth]{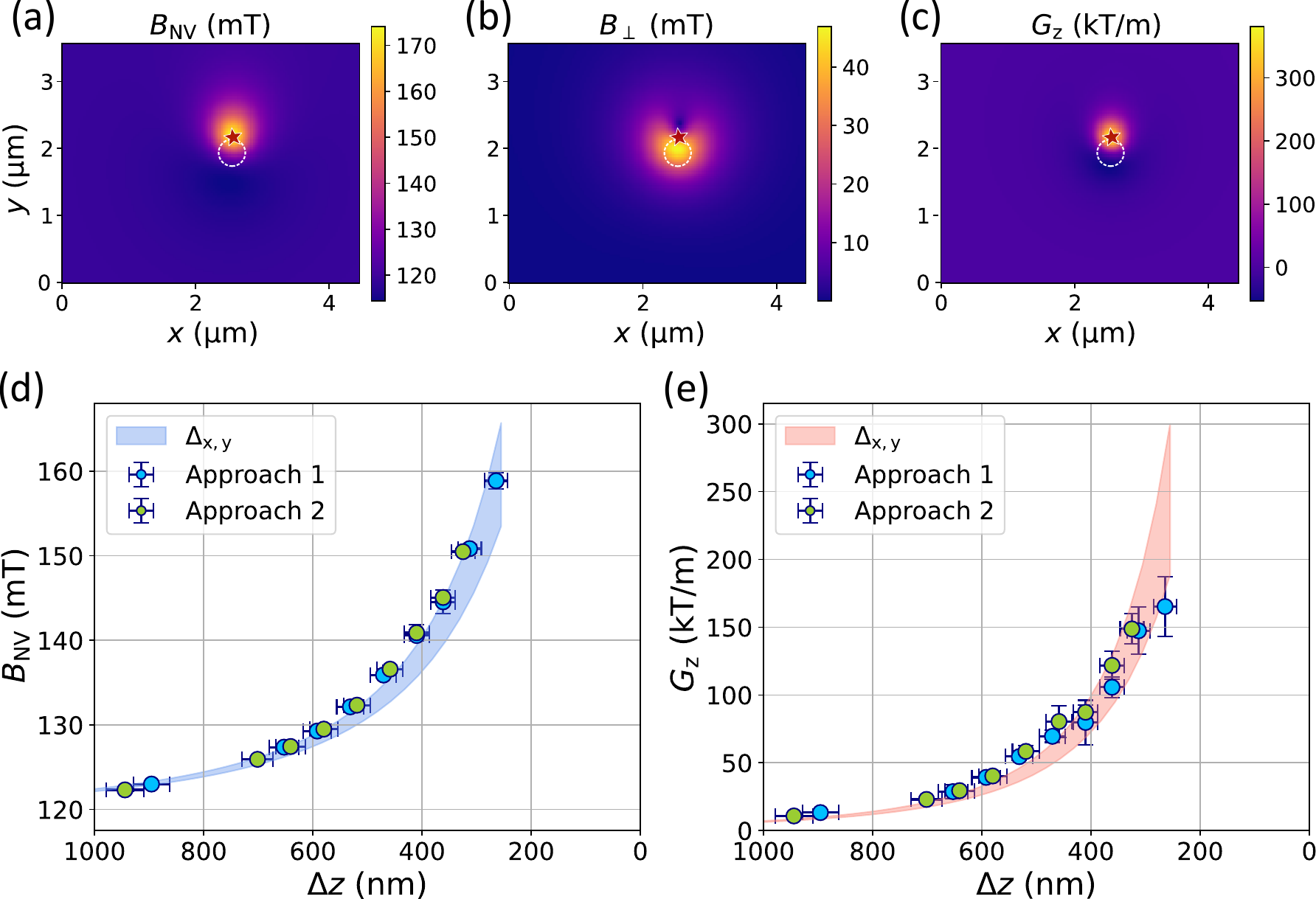}
    \caption{\small{
    Optimal approach region from simulations for measurement of the vertical gradient.
    (a-c) Simulation maps of $\Bnv$, $\Boff$ and $\Bnvgrad$ with  the same $x,y$-axes as \cref{dipolefit}(a), but at a height $h=\qty{200}{\nano\meter}$ above the nanomagnet.
    The white dotted circles mark the position and width of the simulated nanomagnet, the red star highlight the approach location.
    (d) Plots of the measured magnetic field $\Bnv(z)+\Bb$  and (e) the gradient $\Bnvgrad(z)$ from two approaches as a function of the distance to the nanomagnet $\Delta z$.  
    The zero points on the abscissa mark the AFM contact with the surface which was measured before each approach. 
    $\Delta_\mathrm{x} = \qty{111}{\nano\meter}$ and $\Delta_\mathrm{y} = \qty{108}{\nano\meter}$ correspond to the width of the vertically approached region.
    A description of the error bars are in SI \cref{suppl:gradient measurement}.
  }}
  \label{gradient}
\end{figure*}
Foreseeing a future implementation of the experiment in which the magnetic structure is deposited on a high-$Q $ resonator \cite{Tsaturyan2017}, enabling coupling between the out-of-plane motion and the NV center, we are interested in a direct measurement of the magnetic gradient along the $z$ axis.

Under the applied $\Bb$ we measured the gradient $\Bnvgrad$ directly by the Zeeman shift from ODMR spectra as the NV center was scanned through the stray field near the nanomagnet.
Guided by both the dipole model fit and the simulation, we identified the region of highest projected field gradient $\Bnvgrad$, while minimizing magnetic field $\Boff$ that is transverse to $\nnv$, ensuring that the conditions of the \cref{eq:Zeeman interaction} remains valid.

As shown in \cref{gradient}(a-c) we find this region---which is governed by the relative angles between NV and magnetic dipole orientation---towards the edge of the nanomagnet.
There, we approached the nanomagnet in the $z$ direction and acquired ODMR spectra of the lower energy spin resonance at multiple positions to extract the projected magnetic field $\Bnv(z)$ as a function of distance to the nanomagnet surface $\Delta z$.

In \cref{gradient}(d), we present the measured stray fields $\Bnv(z)$ for the two approach trajectories exhibiting the highest field gradients, recorded at a bias field $\Bb \sim \SI{120}{\milli\tesla}$.
The corresponding field gradients $\Bnvgrad(z)$ are shown in \cref{gradient}(e), with maximum measured values of $\Bnv = \SI{40}{mT}+\Bb$ and $\Bnvgrad = \SI{170}{\kilo\tesla\per\meter}$ obtained $\sim\qty{250}{nm}$ above the nanomagnet. 
Acquiring ODMR spectra closer to the surface is hindered by the reduced ODMR signal contrast, which lowers the signal-to-noise ratio below unity (see SI \cref{suppl:gradient measurement}). We attribute this to mechanical drifts in the setup in the high-gradient field, primarily caused by microwave-induced heating.

\begin{figure*}
\center
  \includegraphics[width=0.8\linewidth]{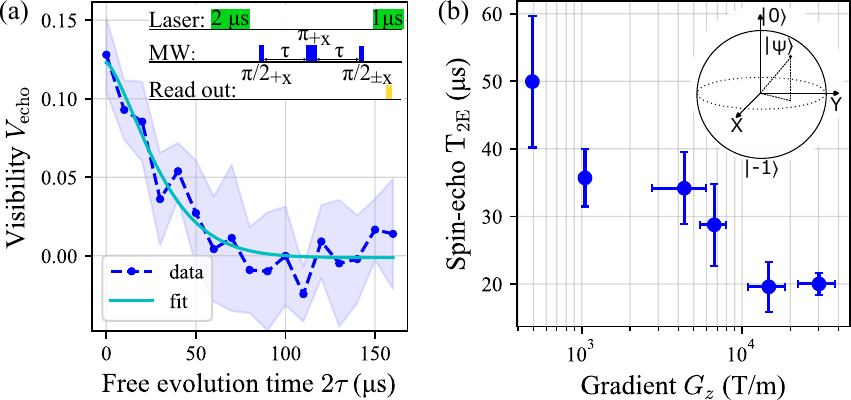}
  \caption{\small{Controlling the NV center's spin in a gradient field.
  (a) $T_\mathrm{2E}$ measurement in the gradient field of the nanomagnet.
  The inset shows the used pulse scheme. 
  Visibility is calculated following \cite{bar2013solid} from two separate spin echo measurements where the last $\pi/2$ pulse is either applied along the $+X$ or $-X$ axis. This measurement was done at $G_z =4.4(1.6)\,\si{\kilo\tesla\per\meter}$  and gives $T_\mathrm{2E} = 34.0(5.3)\,\si{\micro\second}$.
  The shaded area shows the standard deviation. 
  (b) Multiple $T_\mathrm{2E}$ measurements at different positions in the gradient field $G_z$ of a nanomagnet. 
  The measurements were done at distances between approximately \SI{1587}{\nano\metre} and \SI{540}{\nano\metre}.
  The Bloch sphere in the inset defines the axis of the rotation. 
  }}
  \label{t2vsgrad}
\end{figure*}

After characterizing the DC behavior of the spin-mechanical system, we study the dynamics of the NV center using coherent spin control.
We measured the spin decay time under spin-echo sequence, $T_\mathrm{2E}$, while the NV probe was positioned above the nanomagnet (\cref{t2vsgrad}(a)).
The $\pi$-pulse time was first characterized with a Rabi measurement, 
from which we extract the $\pi$-pulse durations of $\sim\SI{55}{\nano\second}$. 
The spin-echo scheme shown in the inset of \cref{t2vsgrad}(a) was applied in two different variations, where the rotation axis of the final $\pi/2$-pulse is either along the $X$ or the $-X$ axis as defined in the Bloch sphere in the inset of \cref{t2vsgrad}(b).
Defining $I_X$ and $I_{-X}$ as the measured photoluminescence rate at the end of a spin-echo sequence for the two readout configurations, we compute the visibility $V_{\mathrm{echo}} = (I_{ X} - I_{ -X})/(I_{X} + I_{-X})$, which rejects the common-mode noise \cite{bar2013solid}. 
The visibility $V_{\mathrm{echo}}$ is then fitted with an exponential decay to extract $T_\mathrm{2E}$. 
This measurement was repeated at different positions in the gradient field $\Bnvgrad$.
In \cref{t2vsgrad}(b) the dependency of $T_\mathrm{2E}$ on $\Bnvgrad$ is shown.
The shift due to instabilities during $T_\mathrm{2E}$ measurements are indicated by horizontal error-bars. 
Applying a MW drive in the vicinity of the NV and the nanomagnet causes heating leading to a shift of ODMR resonance. 
Therefore it is not possible to apply accurate on-resonance pulses at gradients higher than $\SI{30}{kT/m}$.
The vertical error-bars are extracted from the fit of $T_\mathrm{2E}$.

Starting from a value $T_\mathrm{2E} =\SI{56(6)}{\micro\second}$ for a measurement far away from the nanomagnet, $T_\mathrm{2E}$ in \cref{t2vsgrad}(b) shows a significant decrease in value towards higher gradients. 
This trend can be attributed to either magnetic field noise from the magnetic nano-structure, or thermal drifts of the system, which translates into fluctuations in the frequency of the NV center causing dephasing \cite{Lee2017}.

\begin{figure*}
\center
  \includegraphics[width=0.5\linewidth]{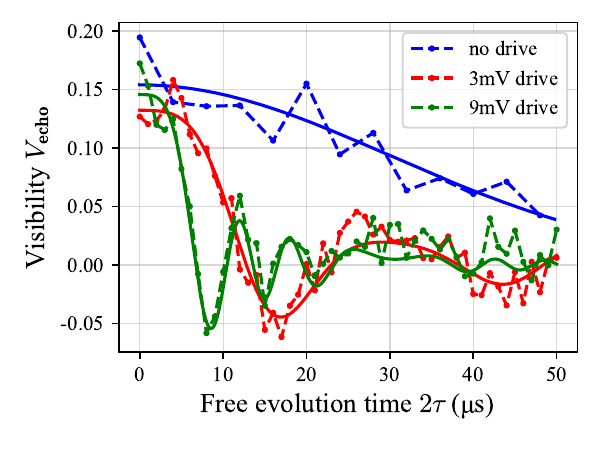}
  \caption{\small{Impact of mechanical oscillation of the tuning fork on the NV center. Blue curve shows a spin-echo measurement for an undriven tuning fork.
  The red and green curves show the measurement with a driven tuning fork at $\Omega_\mathrm{m} \approx \SI{32}{\kilo\hertz}$ with $\SI{3}{\milli\volt}$ and $\SI{9}{\milli\volt}$ driving amplitude, receptively.
  Data (dotted lines) can be fitted with a Bessel function of 0th order (full line).
  }}
  \label{drive}
\end{figure*}

To show that the presented setup is suitable for performing a spin-mechanical experiment, we further measure the effect of the tuning fork mechanical oscillation on the spin dynamics. 
As the tuning fork is driven at its resonance frequency and vibrates along the $x$ direction,
we measure a series of unsynchronized spin-echo sequences (\cref{drive}) with the NV center
moving in a gradient $G_x = \partial \Bnv/\partial x\simeq \SI{540}{\tesla\per\meter}$.

When the tuning fork is not driven, we see undisturbed decay of the visibility with a $T_\mathrm{2E} = \SI{42(3)}{\micro \second}$ (blue curve).
In contrast to that, a driven tuning fork results in a modulated decay (red and green curves). 
It can be fitted to a zeroth order Bessel function, \cite{Fung2024}
\begin{equation}
    V_{\mathrm{echo}} = \frac{C J_0\bigg(\frac{4\pi \gamma_{\mathrm{NV}} x_0 G_x}{\Omega_\mathbf{m}} (\cos(\Omega_\mathrm{m}\tau)-1)\bigg)}{2-C} \exp\left[-\left(\frac{2\tau}{T_\mathrm{2E}}\right)^n\right]\,,
\end{equation}
clearly indicating spin mechanical coupling,
where $C$ is the contrast of our measurement and $x_0$ is the motional amplitude of the NV center (see SI \cref{suppl:pulses}).

The oscillation amplitude values extracted from the fit are $x_0= \SI{3(1)}{\nano \meter}$ (red curve) for a piezo drive amplitude of the tuning fork of \SI{3}{\milli\volt}, and $x_0= \SI{11(1)}{\nano \meter}$ (green curve) for a drive amplitude of \SI{9}{\milli\volt}.
In the two configurations, the mechanical motion maps into an oscillating magnetic field with amplitude  $\SI{2.0(6)}{\micro\tesla}$ and $ \SI{6.1(8)}{\micro\tesla}$ at the position of the NV center, respectively.
\\


In conclusion, we have demonstrated the generation of strong magnetic field gradients—up to $\SI{170}{\kilo\tesla\per\metre}$—using a soft cobalt nanostructure integrated into a spin-mechanics platform based on a scanning NV microscopy configuration.
Spin coherence measurements performed in magnetic field gradients 
up to $\SI{25}{\kilo\tesla\per\metre}$ revealed a coherence 
time $T_\mathrm{2E}$ of at least $\SI{20}{\micro\second}$. 
Considering that the experiments were 
conducted under open-loop mechanical conditions, we expect that,
with cryogenic operation and active stabilization, spin manipulation at even
higher field gradients could be achievable. Moreover, we observed a clear 
influence of the tuning fork motion on the NV spin coherence, indicating the 
presence of measurable spin-mechanical coupling in our setup. 

Crucially, the non-invasive nature of the nanomagnet deposition via FEBID 
allowed us to successfully  decorate a silicon nitride membrane \cite{Catalini2020} with the
magnetic structure (see SI \cref{suppl:nanomagnet_deposition}). This opens the path for further studies, including the 
investigation of potential changes in the mechanical quality factor resulting 
from the deposition---an essential step for future quantum spin-mechanics 
applications. 
If the membranes degradation is negligible, and we consider a typical resonance frequency of  $\SI{1.5}{\mega\hertz}$, an effective mass of $\SI{2}{\nano\gram}$ and an ultra-high quality factor\cite{Rossi2018} $Q= 10^9$  at $\SI{4}{\kelvin}$, then with a
spin coherence time of $\SI{1}{\milli\second} $  and a magnetic gradient
of $\SI{170}{\kilo\tesla\per\meter}$
we obtain a single-phonon coupling strength $g_0/2 \pi= \SI{8}{\hertz}$.
This corresponds to a cooperativity $C=  5 \times 10^{-3}$, which is four orders magnitude higher
than current state-of-the-art setups \cite{Fung2024}.
Remarkably, the force sensitivity of such a device---limited by the thermal contribution  $S_\mathrm{th}= 4 k_\mathrm{B} T m \Omega/Q = \SI{4}{\atto\newton^{2}\per\hertz}$ (with $k_\mathrm{B}$ the Boltzmann constant)---would be comparable to the amplitude of the force produced by the oscillating spin $F= \mu_\mathrm{B} G= \SI{1.6}{\atto\newton}$, thus making it possible to detect the force exerted by the single spin within a few seconds averaging time, during which the spin can be re-polarized as necessary. 

Reaching the milestone of $C\sim 1$ requires technical advances, and appears feasible with the optimistic value\cite{bar2013solid} of $T_2=\SI{10}{\milli\second}$,
and a gradient of $\SI{1}{\mega\tesla\per\meter}$, predicted at a feasible distance of $\qty{150}{nm}$ from the magnetic structure at saturation.

Furthermore, the demonstrated ability to grow nanomagnets directly on membranes
provides a promising platform for implementing quantum sensing experiments. 
While membranes have recently been integrated into scanning force microscopy 
setups \cite{Halg2021,Gisler2024} and adopted in sample-on-resonator 
configurations for magnetic resonance force detection \cite{Scozzaro2016}, we 
foresee their use in magnetic resonance force microscopy experiments in the practical "magnet-on-resonator" 
configuration \cite{Longenecker2012}, as well as for advancing nanoscale magnetic 
resonance imaging techniques \cite{Eichler2022}.

Finally, the high degree of control offered by the FEBID technique enables the
tailoring of nanomagnet geometry to suit specific experimental requirements of the magnetic field, 
making it a versatile tool for future developments in hybrid quantum systems.

\appendix
\label{suppl}
\section{Experimental Setup}
\subsection{Microwave Antenna Fabrication}
Microwave striplines for the experiments are fabricated on high-resistivity (\qty{> 20000}{\ohm\centi\metre}) 4-inch silicon wafers, polished on one side and \qty{500\pm25}{\micro\metre} thick.
A \qty{>1.5}{\micro\metre} layer of AZ 701 MIR photoresist is spin-coated, soft-baked, exposed with a maskless aligner, post-baked, and developed in TMAH.
After surface cleaning with an ion gun in an electron beam evaporator, a \qty{5}{\nano\metre} chromium adhesion layer and \qty{100}{\nano\metre} gold layer are deposited.
Lift-off is performed in 1165 remover, followed by cleaning in IPA, deionized water, and spin drying.
To protect the structures, another photoresist layer is applied and baked and the wafer is diced with a diamond saw.
Individual chips are cleaned in acetone and IPA and wirebonded to the printed circuit board before use.
\begin{figure}[h!]
  \includegraphics[width=0.6\textwidth]{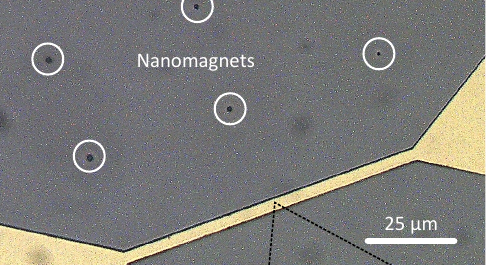}
  \caption{\small{Micrograph of the tapered down region of the MW stripline. Individual nanomagnets are deposited near the stripline as highlighted.}}
  \label{fig:MW stripline}
\end{figure}

\subsection{FEBID Growth of Nanomagnet} \label{suppl:nanomagnet_deposition} 
\begin{figure*}[h!]
  \includegraphics[width=0.4\textwidth]{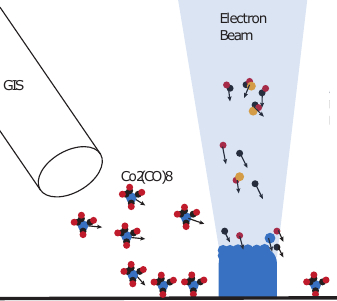}
  \caption{\small{Schematic of a FEBID process where a rectangular beam pattern is employed to create a structure from precursor molecules injected by the gas injection system (GIS). 
  }}
  \label{fab1}
\end{figure*}

\begin{figure}[h!]
  \includegraphics[width=0.6\textwidth]{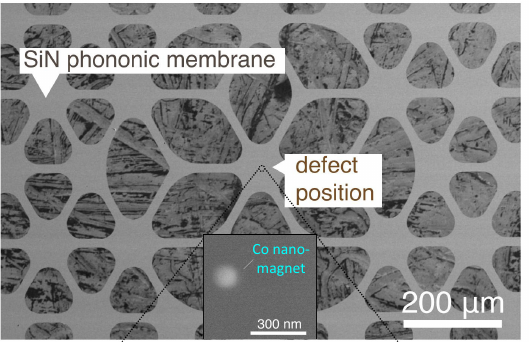}
  \caption{\small{ (a) Top view of a SiN membrane, indicating the position of deposited area.
  Inset: an SEM image of a Co nanomagnet grown on a SiN membrane using a rectangular beam pattern, resulting in a circular-shape structure with a diameter of $\SI{100}{nm}$. }}
  \label{fab2}
\end{figure}

The nanomagnets are positioned at distances ranging from $\qtyrange{30}{210}{\micro\meter}$ from the stripline and are laterally separated from each other by $\qty{60}{\micro\meter}$ in a grid pattern—sufficient to prevent mutual interaction via stray fields.
The deposition starts by releasing the precursor molecule $\mathrm{Co}_2(\mathrm{CO})_8$ in the vacuum chamber through a gas injection system (GIS, see \cref{fab1}). 
The desired geometry is defined as a circular pattern in the computer-aided interface, where the irradiated precursor molecule adsorbs on the surface as the focused electron beam follows the circular raster pattern from the center outward. Initially, we deposited arrays of 4 by 4 circular geometry nanomagnets with defined pattern diameters and heights: $\SI{250}{nm}$, $\SI{200}{nm}$, $\SI{150}{nm}$, $\SI{100}{nm}$ and $\SI{50}{nm}$. 
With the goal in mind to deposit Co nanomagnets on thin, insulating SiN membranes, which exhibit high charging when exposed to an electron beam \cite{De2011}, we opt for the following FEBID parameters: acceleration voltage of $\SI{5}{kV}$, beam current of $\SI{100}{pA}$, dwell time of $\SI{1}{\mu s}$, and precursor flux corresponding to a vacuum pressure varying in the range of $(1.72 - 2.78) \times 10^{-6}\ \mathrm{mbar}$. 
We note the deposited structures show a systematic offset from the defined deposit diameter, ranging between $\SI{20}{nm}$ and $\SI{100}{nm}$. 
The offset may come from the charging effect which is commonly referred to as the halo, and it is produced through a dissociation of the adsorbed precursor by a cascade of secondary electrons generated through backscattering of electrons off the substrate and the grown deposit \cite{doi:10.1021/acsanm.3c05470, Magen2021}. 
Further, we note that the defined pattern influences the charging effect.
In a circularly defined pattern the beam irradiation time per area is maximum in the center of the structure, causing higher charging over time, while for a raster-scanned rectangular pattern the deposition is more uniform.
For our purpose of growing test nanomagnets on silicon chip this effect is seen as an increase in the diameter. 
Conversely, in the growth of nanomagnets on silicon membranes (\cref{fab2}), where the substrate is an insulating material $\mathrm{SiN}$, and has a thickness of $\SI{50}{nm}$, we used a rectangular pattern that combined with the scattering of electrons at the edges results in a circular-shaped structure.

\subsection{External Bias Field}
A $\SI{4}{\centi\meter}$ diameter spherical NdFeB magnet (Supermagnete, K-40-C) with $\Br = \qtyrange{1.26}{1.29}{T}$ residual field density is used.
The magnet is inserted into a 3D printed cage which secures it in place while allowing manual rotation.
Combined with translation stages, a rotator and a goniometer, the magnet can be freely aligned.
The alignment is performed by centering the frequency splitting of the spin resonances $\nu_\mathrm{res}^\pm$ from \cref{eq:Zeeman interaction} symmetrically around $\nu_0$.
According to \cite{Camacho2013} we calculate the norm $|\grad\Bb(r,\theta)| \simeq \qty[per-mode = symbol]{10}{\tesla\per\meter}$ of the magnetic field produced by a spherical permanent magnet at a distance $r(\Bb = \qty{120}{\milli\tesla},\theta = \qty{30}{\degree})$ that matches the scanning plane position.
A Comsol simulation of such a permanent magnet verifies this estimation.
With this at hand, the variation of $\Delta\Bb$ across the scanned map shown in \cref{dipolefit} (a) is expected to be $\qty{30}{\micro\tesla}$, which low compared variation of the nanomagnetic stray field of a couple of mT.

\subsection{Characterization of the Nanopositioners}

\subsubsection*{\texorpdfstring{$z$}{z} - Characterization} \label{suppl:z characterization}
Long-lasting PL oscillations are observable when moving the NV probes in the $z$ direction in the vicinity ($\sim \SI{5}{\micro\meter}$) to the sample surface, that correspond to fluctuations in the excitation of the NV due to self-interference of the excitation laser with its back-reflection at the sample surface \cite{Israelsen2014}. 
These fringes occur with the laser wavelength $\lambda_\mathrm{L} = \qty{515}{nm}$ and by comparing them with measured fringe period, we can estimate the voltage-to-distance ratio $\Cz$ of the $z$-scanner.
Analyzing 47  PL traces with such oscillations, acquired equivalently with the same sample and NV probe over multiple weeks, we did not observe a systematic change in $\Cz$ over time that exceeds statistical fluctuations.

An angle $\alpha$ between the reflection plane and the normal plane (see \cref{magnet NV angles}(a)) increases the interference fringe spacing to \begin{equation}
    \Cgen(\alpha) = \frac{2\Cz}{1 + \cos(2\alpha)}\,.
    \label{gen_CF}
\end{equation}
We probe the angle $\alpha$ by measuring individual AFM contact points on the sample.
In a fixed-point numerical iteration we find the fixed points
\begin{equation}
    \Cgen = \qty[per-mode = symbol]{1214(39)}{\nano\metre\per\volt}
    \,,\qquad
    \alpha = \qty{19.42(24)}{\degree}\,.
    \label{eq:Czgen final result}
\end{equation}
We verify $\Cgen$ by comparing the topographic heights of the nanomagnet measured with the presented scanning NV platform and with a commercial calibrated AFM (Bruker, Dimension Icon PT) and find good agreement using $\Cgen$ as the conversion factor for the scanner voltages.
All shown $z$-scanning distances in the main text are obtained from $\Cgen$ including a NV depth inside the diamond probe apex of $\dnv = \qty{70(20)}{nm}$.

\subsubsection*{\texorpdfstring{$x,y$}{x,y} - Characterization}
We use a checker pattern grid (Anfatec, UMG01B) with well defined trench separations $(\Delta x_\mathrm{p},\Delta y_\mathrm{p})$ to characterize the $x,y$-scanning axes of the nanopositioner under the assumption that the two scanning axes are independent of each other.
By measuring the voltage position $V_\mathrm{x,p}$ of the pillars we can approximate the local piezo gain
\begin{equation}
    g_\mathrm{x}(V_\mathrm{x}) = \dv{x}{V_\mathrm{x}} \simeq \frac{\Delta x_\mathrm{p}}{\Delta V_\mathrm{x,p}}\,,
\end{equation}
by fitting a polynomial of second order to the measured distribution of pillar separations $\Delta x_\mathrm{p}/\Delta V_\mathrm{x,p}$.
Integrating the piezo gain $g(V_\mathrm{x})$ in the scanned voltage region $[V_\mathrm{x}^1, V_\mathrm{x}^2]$ yields the actual scanned spatial distance $\Delta x$
\begin{equation}
    \Delta x = \int_{V_\mathrm{x}^1}^{V_\mathrm{x}^2} g_\mathrm{x}(V_\mathrm{x}) \,dV_\mathrm{x} \,.
\end{equation}
For the scanning $y$-axis, $g_\mathrm{y}(V_\mathrm{y})$ and $\Delta y$ are calculated analogously.

\section{Dipole Fit} \label{suppl: dipole fit}
\subsubsection*{Individual Dipole Fit}
\begin{figure*}[t!]
  \centering
  \includegraphics[width=0.85\textwidth]{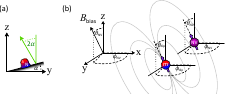}
  \caption{\small{Geometric overview.
  (a) Illustration of the sample tilt. Then green arrow is the back-reflected plane wave from the excitation laser
  (b) Summary of the NV and magnetic dipole angles.
  }}
  \label{magnet NV angles}
\end{figure*}

We use the model of a magnetic dipole $\Bdip$ projected on to the axis of the NV center $\nnv(\phnv,\thnv)$, 
\begin{equation}
\begin{split}
 \Bnv(\vb{R}) &= \Bdip(\vb{R})\cdot\nnv +\Bb \\
  & = \frac{\mu_0}{4\pi }\left( \frac{3\vb{R}(\vecmum\cdot\vb{R})}{R^5 } - \frac{\vecmum}{R^3} \right)\cdot \nnv + \Bb\,,
 \end{split}
\end{equation}
to fit the measured stray field of the nanomagnet in the main text (\cref{dipolefit}).
We are scanning in planes at heights from $\qtyrange{960}{1460}{\nano\metre}$ above the nanomagnet.
This includes the constant offset from the NV-to-sample distance which we measured independently in AFM contact to be $\qty{70(20)}{nm}$.
We define the scanning plane as  $\vb{R}(x,y) = (x-x_0, y-y_0, 0-z_0)$, with dipole position $(x_0, y_0, z_0)$, dipole magnetic moment $\vecmum$ and vacuum magnetic permeability $\mu_0$.
We approximate $\left(\nnv(\phnv,\thnv) \parallel \vecmum (\phi_\mathrm{dip},\theta_\mathrm{dip})\right)$ (see \cref{magnet NV angles}(b)), which reduces the fit parameter space to $\{ x_0,y_0,z_0,\phnv, \thnv,\mum,\Bb\}$.

\subsubsection*{Simultaneous Dipole Fit}
Incorporating the known height steps $\zstep$ in a simultaneous fit across the scanned maps of a set enforces consistency with the vertical decay trend of a dipolar field, $\Bnv \propto \mum/\rzz^3$, thereby naturally constraining the strongly correlated  $\rzz$ and the magnetic moment $\mum$ parameters.
The individual scanned maps can spatially vary in the $xy$-plane from each other.
Pre-computation of the $x_0,y_0$ dipole positions of each map with the individual dipole fit allows further constrains.
The global fit parameter space is then reduced to a $\{z_0,\phnv, \thnv,\mum,\Bb\}$.

\section{Micromagnetic Simulation of the Nanomagnet} 
\label{suppl:micromatnetic simulation}
\begin{figure*}[t!]
  \centering
  \includegraphics[width=0.65\textwidth]{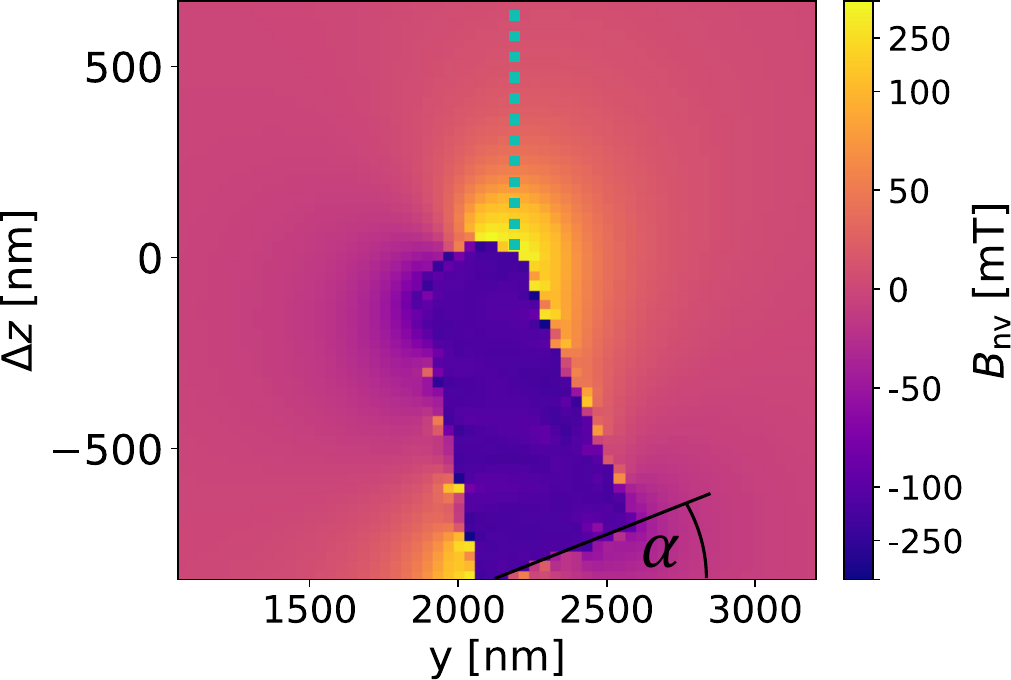}
  \caption{\small{Cross section of the simulated nanomagnet in the scanning frame scanning frame denoted in \cref{setup}(b). 
  The dashed line indicates the center of the approach region for the gradients measured in the main text with width $\Delta_\mathrm{x,y}$.
  The angle $\alpha$ resembles the plane tilt.
  The colorbar is linear within $\qty{\pm100}{mT}$ and logarithmic beyond that.
  }}
  \label{Simulation Cross Section}
\end{figure*}
We use the finite-difference micromagnetic simulation software MuMax3 \cite{Vansteenkiste2014,Exl2014} to simulate the stray field of the FEBID grown nanomagnet using the parameters reported in \cite{doi:10.1021/acsanm.3c05470, Cheenikundil2021}: exchange stiffness constant $A_\mathrm{ex} = \qty{14e-12}{\joule\per\meter}$, zero magnetocrystalline anisotropy and a saturation magnetization $M_\mathrm{sat} = \qty{1.26}{\ampere\per\meter}$ which is $\qty{90}{\percent}$ of the bulk value based on comparable literature on non-annealed structures \cite{doi:10.1021/acsanm.3c05470, DeTeresa2016}.

The geometry of the simulated nanomagnet is based on the AFM topography acquired with a commercial AFM (Bruker, Dimension Icon PT).
The simulation voxel size is $(\qty{36.83}{}, \qty{30}{}, \qty{25}{})\,$nm with a grid size of $(120,100,110) \,$cells.
We include an external bias field parallel to $\nnv$ with angles given by the dipole fit \cref{dipolefit} (a) from the main text.
To match the sample plane tilt $\alpha$ from \cref{suppl:z characterization}, we rotate the nanomagnet accordingly.
A cross section of the simulated nanomagnet is shown in \cref{Simulation Cross Section}.
In the hysteresis the external bias field is varied from $\qty{0}{mT}\rightarrow\qty{750}{mT}\rightarrow\qty{-750}{mT}\rightarrow\qty{750}{mT}$ in increments of $\qty{25}{mT}$.

\section{Gradient Measurement} \label{suppl:gradient measurement}
To estimate the stray field $\Bnv$ we use a Lorentzian model\cite{Balasubramanian2008} to evaluate the ODMR spectra. 
As mentioned in the main text, every data point in \cref{gradient} is evaluated from the average spectrum of three ODMR spectra taken consecutively.
A first Lorentzian fit is used to define an ODMR window as twice the FWHM around the center frequency $\nu_0$ of the ODMR dip.
We assume gaussian noise as our base noise and calculate the standard deviation $\mad = k_\mathrm{gauss}\cdot\mathrm{MAD}$ of the PL outside the defined ODMR window, where MAD is the median absolute deviation and $k_\mathrm{gauss}$ a scale factor for normally distributed noise.
We have observed that our data acquisition card (National Instruments X Series) can induce single data point spikes in the measured PL. We attribute these outliers to a buffer overflow of the card.
To exclude these outliers in a second fitting step, data points inside as well as outside the window that deviate by more than $\qty{3}{\mad}$ to the first Lorentzian fit are excluded.
From the outlier-free second Lorentzian fit we estimate stray field of the nanomagnet via \cref{eq:Zeeman interaction}.
Only ODMR measurements are considered that fulfill the two conditions:
\begin{itemize}
  \item The frequency window of the full spectrum $\Delta\nu > 4\Gamma$.
  \item The ODMR contrast $C > \qty{1}{\mad}$.
\end{itemize}
We argue that the first condition ensures to have enough statistical data to estimate a tangible MAD and to differentiate the ODMR signal from slow PL drifts. 
The second condition ensures $\mathrm{SNR} > 1$.
We give $\Gamma$ as a conservative uncertainty for the stray field estimation $\Delta\Bnv$.
The data acquisition time was set to $\qty{47}{ms}$ per MHz bandwidth resulting in a $\qtyrange{10}{20}{s}$ of continuous wave MW emission for each data point - depending on the frequency range of the ODMR spectrum.

To probe and locate the region for the approaches shown in \cref{gradient}(d,e), we vertically approach the surface along the indicated line in \cref{Simulation Cross Section} using the nanopositioners while monitoring the PL.

\subsubsection*{Error Bars}
The gradient is calculated via the gradient function of the Python package Numpy which uses central differences and forward/backwards difference for the boundary points.
Therefore, we propagate $\Delta\Bnv$ into the vertical error bars $\Delta\Bnvgrad$ of \cref{gradient} for interior and boundary points separately.
The error $\Delta\Cgen$ from \cref{eq:Czgen final result} translates into the horizontal error bars quadratically combined with the $\Delta\dnv$.

\section{Pulse measurements}  \label{suppl:pulses}
\subsection*{\texorpdfstring{$T_\mathrm{2E}$ vs gradient}{T2E measurement}} \label{sec:T2E measurement}
\begin{figure*}[h!]
  \includegraphics[width=0.7\textwidth]{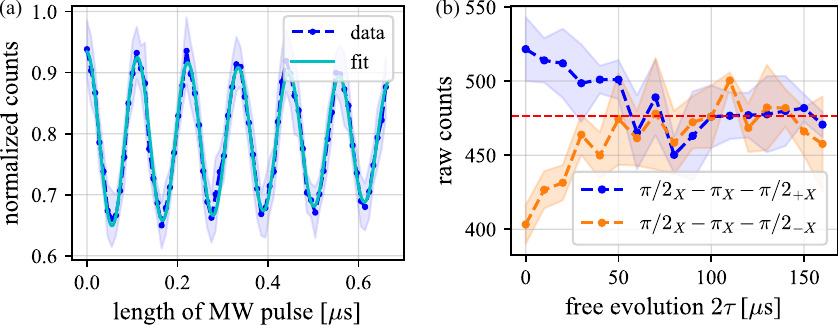}
  \caption{\small{(a)  Rabi oscillations used to calibrate the $\pi$- and $\pi/2$-pulses. Usual $\pi$- pulse times of ca. $\SI{55}{ns}$ are extracted from the fit. 
  The shaded area represents the standard deviation observed across multiple runs of the measurement. 
  (b) Spin-echo measurements, for calculating visibility in the main text.
  The shaded area represents the standard deviation observed across multiple runs of the measurement. The red line indicates the baseline of the background counts. 
  }}
  \label{t2e_sup}
\end{figure*}

At every position in the gradient field the $\pi$- pulse time was first characterized with a standard Rabi measurement \cref{t2e_sup}(a).
A $\pi$-pulse is a resonant microwave pulse that inverts the population of the two level system. When applied around the $X$-axis, it can effectively flip the state from $\ket{0}$ to $\ket{1}$.

Fitting the Rabi oscillations with 
\begin{equation}
A \sin(2\pi v_r t + \Phi) \exp(-\Gamma_\mathrm{Rabi} t) + C \,, 
\end{equation} 
we extract $\pi$- pulse durations of $\SI{55}{ns}$ (bright blue line in \cref{t2e_sup}(a)). 
In \cref{t2e_sup} (b) we show the two different $T_\mathrm{2E}$ measurements for the readout pulse around the X and -X axis on the Bloch sphere. 
The two measurement are used to calculate the visibility as mentioned in the main text. 

The horizontal error bars in \cref{t2vsgrad} (main text) are calculated by measuring the gradient before the spin echo measurement and after the measurement was done. 
The first two datapoints in \cref{t2vsgrad} (main text) do not have a horizontal error bars because we did not measure the gradient before and after the spin-echo sequence.

\subsection*{Modeling of the NV signal in presence of a mechanical drive}
The mechanical oscillation, actuated by a piezo shaker driven at the mechanical resonance frequency, can be written as $x(t) = x_0 \cos\left(\Omega_\mathbf{m} t + \phi_0\right)$, where $\Omega_\mathbf{m}, x_0$ and $\phi_0$ are the resonator frequency, amplitude and phase respectively. 

The semi-classical parametric spin-mechanics interaction Hamiltonian is:
\begin{equation}
\mathcal{H}^{\mathrm{int}}/\hbar = \pi \gamma_{\mathrm{NV}}G_\mathrm{x} x(t)\sigma_3, 
\end{equation}
where $G_\mathrm{x} = \partial \Bnv / \partial x$ is the gradient - along the oscillator’s motion direction ($x$) - of the magnetic field component ($\Bnv$) parallel to the NV spin quantization axis.
Here, $\sigma_3$ is the Pauli matrix. 
Starting from the initial $\ket{0}$ state, the system evolves under the following spin echo pulse scheme: $[\pi/2 - \tau - \pi -\tau - \pi/2]$ where $\tau$ is the time between the pulses.
The first $\pi/2$ pulse creates a superposition on the equator of the Bloch sphere which, up to a global phase, can be written as
\begin{equation}
\ket{\Psi_0} = \frac{1}{\sqrt{2}}\left(\ket{0} - i\ket{1}\right)\,,
\end{equation}
in the frame rotating with the spin resonance frequency.
At the end of the free evolution on the equatorial plane, the spin state can be written as 
\begin{equation}
    |\Psi_{2\tau}\rangle = \frac{1}{\sqrt{2}} \left(\ket{0} - ie^{-i\phi}\ket{1} \right)\,,
\end{equation} 
 where the accumulated relative phase reads\cite{Fung2024}
 \begin{equation}
\phi(\tau) = \frac{2 \pi \gamma_{\mathrm{NV}} G_\mathrm{x} x_0}{\Omega_\mathbf{m}}\left[\sin(2\Omega_\mathbf{m}\tau + \phi_0)-2\sin(\Omega_\mathbf{m}\tau + \phi_0) + \sin(\phi_0)\right]\,.  \\
\end{equation}
Lastly, after the final read-out $\pi/2$ pulse, the probability to find the system in the $\ket{0}$ level is\cite{kolkowitz2012coherent} 
\begin{equation}
    P_0 = |\langle \Psi_0 | \Psi_{2\tau} \rangle|^2 = \frac{1}{2} \left(1 \mp \cos(\phi)\right)\,,
\end{equation}
where the sign $\mp$ depends on whether the readout pulse is around the $X$ or the $-X$ axis of the Bloch sphere.
Assuming a uniformly distributed $\phi_0$ during the pulse measurements, we get an average contrast: 
\begin{equation}
    \langle \cos(\phi(\tau))\rangle_{\phi_0} = J_0\left(\frac{4 \pi \gamma_{\mathrm{NV}} G_\mathrm{x} x_0}{\Omega_\mathbf{m}} (\cos(\Omega_\mathbf{m}\tau)-1)\right)\,.
\end{equation}
The visibility $V_{\mathrm{echo}}$, as defined in the main text, reads: 
\begin{equation}
    V_{\mathrm{echo}} = \frac{C J_0\left(\frac{4\pi \gamma_{\mathrm{NV}}G_\mathrm{x} x_0}{\Omega_\mathbf{m}} (\cos(\Omega_\mathbf{m}\tau)-1)\right)}{2-C} \exp\left[-\left(\frac{2 \tau}{T_\mathrm{2E}}\right)^n\right],
\end{equation}
where we have added the exponential decay which accounts for the spin decoherence.

\bibliography{references_ordered}
\end{document}